\documentclass[aps,prd,preprint,nofootinbib,superscriptaddress,showpacs,byrevtex,titlepage,tightenlines]{revtex4}
\usepackage{graphicx} 
\usepackage{epsfig}
\usepackage{dcolumn}  

\graphicspath{{ps}}

\begin{document}

\preprint{\vbox{ 
		 \vskip 5mm
                 \hbox{\hskip-120mm
                       \includegraphics[height=1.7cm]{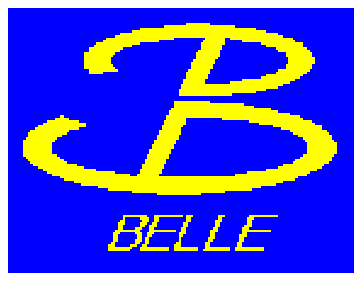}}
		 \vskip -17mm
		 \hbox{BELLE-CONF-0316}
                 \hbox{EPS Parallel Sessions: 3, 10, 12}
                 \hbox{EPS-ID 534}
}}

\title{ \quad\\[0.5cm]  Observation of $B^{\pm} \rightarrow D_{CP}K^{*\pm}$ Decays}

\affiliation{Aomori University, Aomori}
\affiliation{Budker Institute of Nuclear Physics, Novosibirsk}
\affiliation{Chiba University, Chiba}
\affiliation{Chuo University, Tokyo}
\affiliation{University of Cincinnati, Cincinnati, Ohio 45221}
\affiliation{University of Frankfurt, Frankfurt}
\affiliation{Gyeongsang National University, Chinju}
\affiliation{University of Hawaii, Honolulu, Hawaii 96822}
\affiliation{High Energy Accelerator Research Organization (KEK), Tsukuba}
\affiliation{Hiroshima Institute of Technology, Hiroshima}
\affiliation{Institute of High Energy Physics, Chinese Academy of Sciences, Beijing}
\affiliation{Institute of High Energy Physics, Vienna}
\affiliation{Institute for Theoretical and Experimental Physics, Moscow}
\affiliation{J. Stefan Institute, Ljubljana}
\affiliation{Kanagawa University, Yokohama}
\affiliation{Korea University, Seoul}
\affiliation{Kyoto University, Kyoto}
\affiliation{Kyungpook National University, Taegu}
\affiliation{Institut de Physique des Hautes \'Energies, Universit\'e de Lausanne, Lausanne}
\affiliation{University of Ljubljana, Ljubljana}
\affiliation{University of Maribor, Maribor}
\affiliation{University of Melbourne, Victoria}
\affiliation{Nagoya University, Nagoya}
\affiliation{Nara Women's University, Nara}
\affiliation{National Kaohsiung Normal University, Kaohsiung}
\affiliation{National Lien-Ho Institute of Technology, Miao Li}
\affiliation{Department of Physics, National Taiwan University, Taipei}
\affiliation{H. Niewodniczanski Institute of Nuclear Physics, Krakow}
\affiliation{Nihon Dental College, Niigata}
\affiliation{Niigata University, Niigata}
\affiliation{Osaka City University, Osaka}
\affiliation{Osaka University, Osaka}
\affiliation{Panjab University, Chandigarh}
\affiliation{Peking University, Beijing}
\affiliation{Princeton University, Princeton, New Jersey 08545}
\affiliation{RIKEN BNL Research Center, Upton, New York 11973}
\affiliation{Saga University, Saga}
\affiliation{University of Science and Technology of China, Hefei}
\affiliation{Seoul National University, Seoul}
\affiliation{Sungkyunkwan University, Suwon}
\affiliation{University of Sydney, Sydney NSW}
\affiliation{Tata Institute of Fundamental Research, Bombay}
\affiliation{Toho University, Funabashi}
\affiliation{Tohoku Gakuin University, Tagajo}
\affiliation{Tohoku University, Sendai}
\affiliation{Department of Physics, University of Tokyo, Tokyo}
\affiliation{Tokyo Institute of Technology, Tokyo}
\affiliation{Tokyo Metropolitan University, Tokyo}
\affiliation{Tokyo University of Agriculture and Technology, Tokyo}
\affiliation{Toyama National College of Maritime Technology, Toyama}
\affiliation{University of Tsukuba, Tsukuba}
\affiliation{Utkal University, Bhubaneswer}
\affiliation{Virginia Polytechnic Institute and State University, Blacksburg, Virginia 24061}
\affiliation{Yokkaichi University, Yokkaichi}
\affiliation{Yonsei University, Seoul}
  \author{K.~Abe}\affiliation{High Energy Accelerator Research Organization (KEK), Tsukuba} 
  \author{K.~Abe}\affiliation{Tohoku Gakuin University, Tagajo} 
  \author{N.~Abe}\affiliation{Tokyo Institute of Technology, Tokyo} 
  \author{R.~Abe}\affiliation{Niigata University, Niigata} 
  \author{T.~Abe}\affiliation{High Energy Accelerator Research Organization (KEK), Tsukuba} 
  \author{I.~Adachi}\affiliation{High Energy Accelerator Research Organization (KEK), Tsukuba} 
  \author{Byoung~Sup~Ahn}\affiliation{Korea University, Seoul} 
  \author{H.~Aihara}\affiliation{Department of Physics, University of Tokyo, Tokyo} 
  \author{M.~Akatsu}\affiliation{Nagoya University, Nagoya} 
  \author{M.~Asai}\affiliation{Hiroshima Institute of Technology, Hiroshima} 
% ...
\collaboration{The Belle Collaboration}

\begin{abstract}
\noindent

We report preliminary results on the decay $B^{-} \rightarrow D^{0}K^{*-}$, $B^{-} \rightarrow D_{CP}K^{*-}$ and its charge conjugate using a data sample of 95.8 million $B\overline{B}$ pairs recorded at the  $\Upsilon(4S)$ resonance with the Belle detector at the KEKB asymmetric $e^{+}e^{-}$ storage ring. We find the branching fraction for $B^{-} \rightarrow D^{0} K^{*-}$ to be  ${\cal B} = (5.2 \pm 0.5(stat) \pm 0.6(sys)) \times 10^{-4}$ and the partial-rate charge asymmetries for  $B^{-} \rightarrow D_{CP}K^{*-}$ to be ${\cal{A}}_1 = -0.02 \pm 0.33(stat) \pm 0.07(sys)$ and ${\cal{A}}_2 = 0.19 \pm 0.50(stat) \pm 0.04(sys)$ where the indices 1 and 2 represent the CP=+1 and CP=$-$1 eigenstates of the $D^{0}-\bar{D^{0}}$ system, respectively.

\end{abstract}
\maketitle \tighten
{\renewcommand{\thefootnote}{\fnsymbol{footnote}}}
\setcounter{footnote}{0}
\normalsize
	
        The extraction of $\phi_{3}$~\cite{gamma}, an angle in one of the unitarity triangles of the Kobayashi-Maskawa~(KM) quark mixing matrix~\cite{KM}, is a challenging measurement even with modern high luminosity $B$ factories. Recent theoretical work on $B$ meson dynamics has demonstrated the direct accessibility of $\phi_{3}$ using the process $B^{-} \rightarrow DK^{*-}$~\cite{ADS}. If the $D^{0}$ is reconstructed as a CP eigenstate, the $ b \rightarrow c$ and $b \rightarrow u$ processes interfere. This interference leads to direct CP violation as well as a characteristic pattern of branching fractions. However, the branching fractions for $D$ meson decay modes to CP eigenstates are only of order 1~\%. Since CP violation through interference is expected to be small, a large number of $B$ decays is needed to extract $\phi_3$.  Assuming the absence of $D^0 - \bar{D^0}$ mixing, the observables sensitive to CP violation that are used to extract the angle $\phi_3$ are
\begin{eqnarray*}
{\cal{A}}_{1,2} \equiv \frac{{\cal B}(B^- \rightarrow D_{1,2}K^{*-}) - {\cal B}(B^+ \rightarrow D_{1,2}K^{*+}) }{{\cal B}(B^- \rightarrow D_{1,2}K^{*-}) + {\cal B}(B^+ \rightarrow D_{1,2}K^{*+}) }&\\
= \frac{2 r \sin \delta ' \sin \phi_3}{1 + r^2 + 2 r \cos \delta ' \cos \phi_3}~~~~~~~~~~~~~~~~~~~~~~&\\
\delta ' = \left\{
             \begin{array}{ll}
              \delta & \mbox {{\rm  for }$D_1$}\\
              \delta + \pi&  \mbox{{\rm for }$D_2$}\\
             \end{array}
             \right.,~~~~~~~~~~~~~~~~~&
\end{eqnarray*}
where $D_1$ and $D_2$ are CP-even and CP-odd eigenstates of the neutral $D$ meson, $r$ denotes a ratio of amplitudes, $r \equiv |A(B^- \to {\bar{D}^{0}}K^{*-})/A(B^- \to D^0 K^{*-})|$, and $\delta$ is the strong interaction phase difference. The ratio $r$ corresponds to the magnitude of CP asymmetry and is suppressed to the level of $\sim 0.1$ due to the CKM factor~$(\sim 0.4)$ and a color suppression factor~$(\sim 0.25)$. 

	This measurement of the branching fraction for $B^{-} \rightarrow D^{0}K^{*-}$ and CP asymmetries for $B^{-} \rightarrow D_{CP}K^{*-}$ is based on a data sample of 95.8 million $B\overline{B}$ pairs, collected  with the Belle detector at the KEKB asymmetric-energy $e^+e^-$ (3.5 on 8~GeV) collider~\cite{KEKB} operating at the $\Upsilon(4S)$ resonance. At KEKB, the $\Upsilon(4S)$ is produced with a Lorentz boost of $\beta\gamma=0.425$ nearly along the electron beamline.

The Belle detector is a large-solid-angle magnetic spectrometer that consists of a three-layer silicon vertex detector (SVD), a 50-layer central drift chamber (CDC), an array of silica aerogel threshold \v{C}erenkov counters (ACC), a barrel-like arrangement of time-of-flight scintillation counters (TOF), and an electromagnetic calorimeter (ECL) comprised of CsI(Tl) crystals located inside a super-conducting solenoid coil that provides a 1.5~T magnetic field. The iron flux-return located outside the coil is instrumented to detect $K_L^0$ mesons and to identify muons (KLM).  The detector is described in detail elsewhere~\cite{Belle}.
  
	We reconstruct $D^{0}$ mesons in the following decay channels. For the flavor specific mode (denoted by $D_{f}$), we use $D^{0} \rightarrow K^{-}\pi^{+}$, $K^{-}\pi^{+}\pi^{0}$ and $K^{-}\pi^{+}\pi^{-}\pi^{+}$~\cite{CC}. For CP =+1 modes, we use $D_{1} \rightarrow K^{-}K^{+}$ and $\pi^{-}\pi^{+}$ while, for ${\rm CP} =-1$ modes, we use $D_{2} \rightarrow K_{S}^{0}\pi^{0}$, $K_{S}^{0}\phi$, and $K_{S}^{0}\omega$. We reconstruct $K^{*-}$ candidates from $K_{S}^{0}\pi^{-}$ combinations where the $K_{S}^{0}$ candidate is formed from two oppositely charged pions having a vertex displaced from the interaction  point in the direction of the $K_{S}^{0}$ momentum. The $K_{S}^{0}$ candidates are selected in the mass window of $0.492\,\mathrm{~GeV}/c^2 < M(\pi^{+}\pi^{-}) < 0.505\,\mathrm{~GeV}/c^2$. The candidate tracks are then kinematically constrained to its nominal mass value. The $K^{*-}$ is required to have a mass within $\pm75\,\mathrm{~MeV}/c^2$ of its nominal mass value. For the $\pi^{0}$ from the $D^{0}\rightarrow K^{-}\pi^{+}\pi^{0}$ decays, we require the $\pi^{0}$ momentum in the $\Upsilon(4S)$ center-of-mass(c.m.) frame be greater than $0.2\,\mathrm{~GeV}/c$ and the energy of each photon from the $\pi^{0}$ be greater than $30\,\mathrm{~MeV}$.
	
	Well constrained reconstructed tracks that are not identified as electrons or muons are used as charged hadrons. For each charged track, information from the ACC, TOF and specific ionization measurements from the CDC are used to determine a $K/\pi$ likelihood ratio $P(K/\pi)$ = $\it L_{K}/(\it L_{K} + \it L_{\pi})$, where $\it L_{K}$ and $\it L_{\pi}$ are kaon and pion likelihoods. For kaons (pions) from the $D^{0} \rightarrow K^{-}\pi^{+}$ mode we used the particle identification requirement of $P(K/\pi) > 0.4~(<0.7)$. For kaons from the $D^{0} \rightarrow K^{-}K^{+}$ mode we require $P(K/\pi) > 0.7$ while for pions from $D^{0} \rightarrow \pi^{-}\pi^{+}$ mode we require $P(K/\pi) < 0.7$.

	The $\omega$ mesons are reconstructed from $\pi^{+}\pi^{-}\pi^{0}$ combinations in the mass window $0.732\,\mathrm{~GeV}/c^2 < M(\pi^{+}\pi^{-}\pi^{0}) < 0.82\,\mathrm{~GeV}/c^2$ with the charged pion particle identification requirement $P(K/\pi) < 0.8$. To reduce the contribution from the non-resonant background, a helicity angle requirement $|{\rm{cos}}~\theta_{\rm{hel}}| > 0.4$ is applied, where $\theta_{\rm{hel}}$ is the angle between the normal to the $\omega$ decay plane in the $\omega$ rest frame and the $\omega$ momentum in the $D^{0}$  frame. To remove the contribution from $D^{0} \rightarrow K^{*-}\rho^{+}$, we require the $K_{S}^{0}\pi^{-}$ invariant mass to be greater than $75\,\mathrm{~MeV}/c^2$ from the $K^{*-}$ nominal mass.

	The $\phi$ mesons are reconstructed from two oppositely charged kaons in the mass window of $1.008\,\mathrm{~GeV}/c^2 < M(K^{+}K^{-}) < 1.032\,\mathrm{~GeV}/c^2$ with $P(K/\pi) > 0.2$. We also apply the $\phi$ helicity angle cut $|{\rm{cos}}~\theta_{\rm{hel}}| > 0.4$, where $\theta_{\rm{hel}}$ is the angle between one of the $\phi$ daughters in the $\phi$ rest frame and the $\phi$ momentum in the $D^{0}$  frame. The $D^{0}$ candidates are required to have masses within $\pm$2.5$\sigma$ of their nominal masses, where $\sigma$ is the measured mass resolution that ranges from $5\,\mathrm{~MeV}/c^2$ to $18\,\mathrm{~MeV}/c^2$ depending on the decay channel. A $D^{0}$ mass and (wherever possible) vertex constrained fit is then performed on the remaining candidates.
    
	We combine the $D^{0}$ and $K^{*-}$ candidates to form $B$ candidates. The signal is identified by two kinematic variables: the energy difference $\Delta E = E_D + E_{K^{*-}} - E_{\rm{beam}}$ and the beam energy constrained mass $M_{\rm{bc}} = \sqrt{E_{\rm{beam}}^2 - |\vec{p}_D + \vec{p}_{K^{*-}}|^2}$. Here $E_D$ is the energy of the $D^{0}$ candidate, $E_{K^{*-}}$ is the energy of the $K^{*-}$ and $E_{\rm{beam}}$ is the beam energy, and $\vec{p_{D}}$ and $\vec{p}_{K^{*-}}$ are the momenta of the $D^{0}$ and $K^{*-}$ candidates, respectively, all calculated in the c.m. frame.  Event candidates are accepted if they have $5.2\,\mathrm{~GeV}/c^2 < M_{\rm{bc}} < 5.3\,\mathrm{~GeV}/c^2$ and $|\Delta E| < 0.2\,\mathrm{~GeV}$. In case of multiple candidates from a single event, we choose the best candidate on the basis of a $\chi^2$ determined from the differences between the measured and nominal values of $M_D$ and $M_{\rm{K^{*-}}}$. Since $B^{-} \rightarrow D^{0}K^{*-}$ is a pseudoscalar to pseudoscalar-vector decay, the $K^{*-}$ is polarized. We define the $K^{*-}$ helicity angle ${\rm{cos}}~\theta_{\rm{hel}}$ as the angle between one of the $K^{*-}$ decay products in the $K^{*-}$ rest frame and the $K^{*-}$ momentum in the $B$ rest frame. The $K^{*-}$ helicity angle follows a ${\rm{cos}^2}\theta_{\rm{hel}}$ distribution. We require $|{\rm{cos}}~\theta_{\rm{hel}}| > 0.4$.

	To suppress the large combinatorial background from the two-jet-like $e^{+}e^{-} \rightarrow q\bar{q}$ ($q = u$, $d$, $s$ or $c$) continuum processes, variables that characterize the event topology are used. We use a Fisher discriminant $\it F$, constructed from six modified Fox Wolfram moments~\cite{SFW} and $\cos{\theta_{B}}$, the cosine of the angle of the $B$ flight direction with respect to the beam axis, in a single likelihood ratio variable ($LR$) that distinguishes signal from continuum background. We apply a different requirement for each sub-mode based on the expected signal yield and the backgrounds in the $M_{\rm{bc}}$ sideband data. To give an example of the performance of this selection, the $LR > 0.4$ requirement keeps 89.4~\% of the $B^{-} \rightarrow D^{0}[K^{-}K^{+}]K^{*-}$signal while removing 69.2~\% of the continuum background.
	
\begin{figure}[ht]
\begin{center}
 \begin{tabular}{ll}
   \epsfig{file=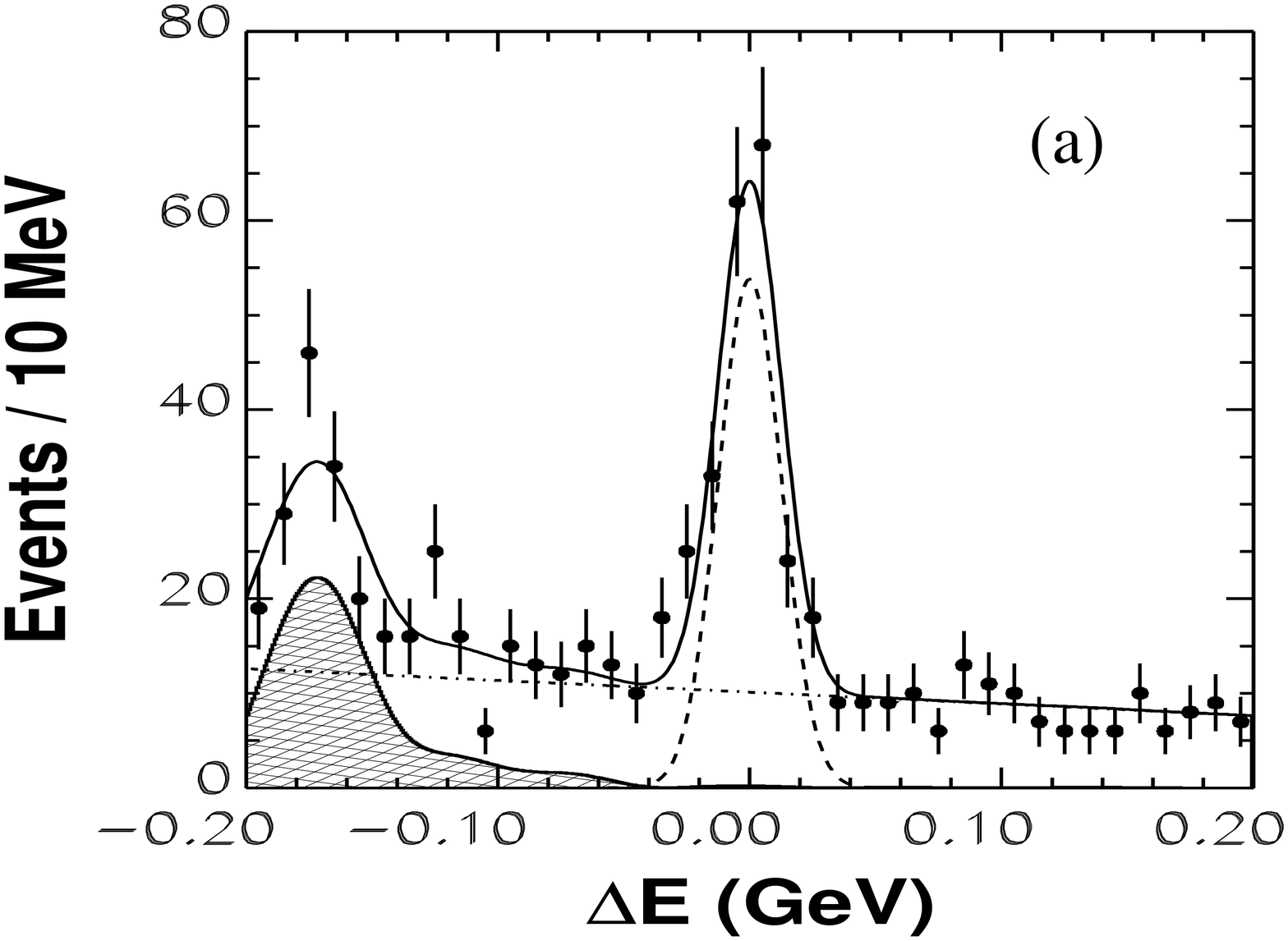,height=3.8cm,width=8cm} &
   \epsfig{file=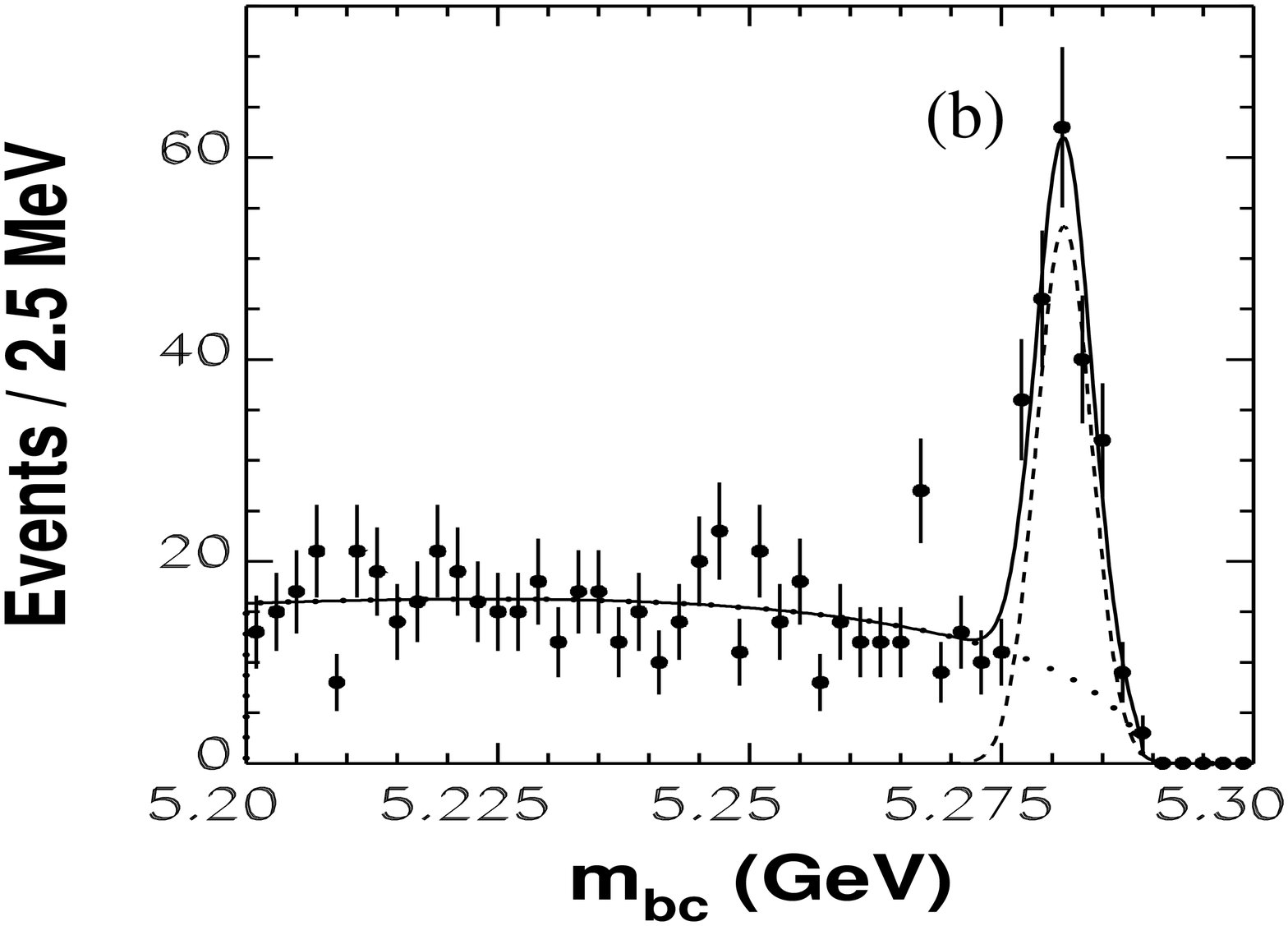,height=3.8cm,width=8.0cm}\\ 
\hskip -0.15cm
   \epsfig{file=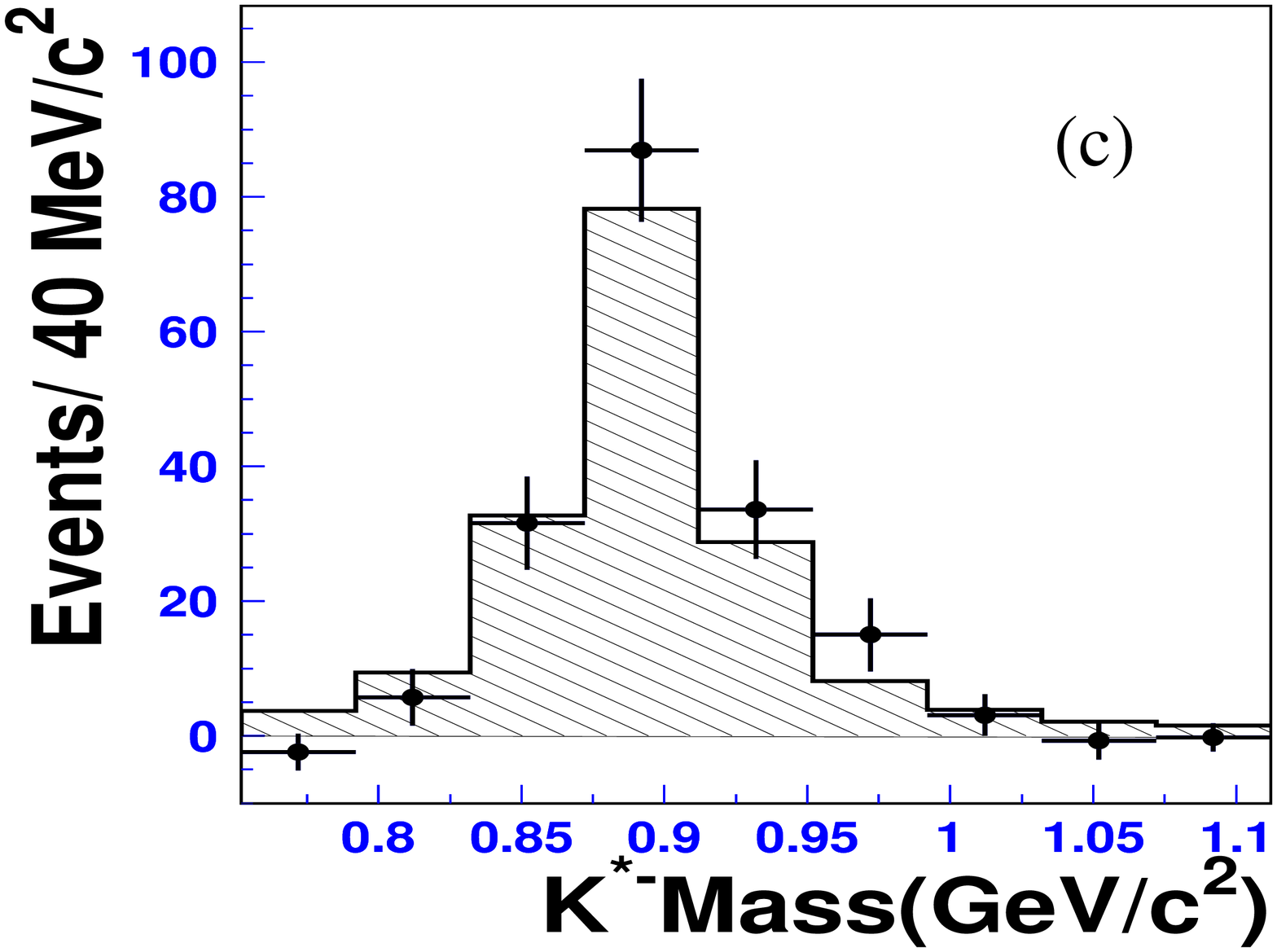,height=3.8cm,width=8.3cm} &
\hskip -0.15cm
   \epsfig{file=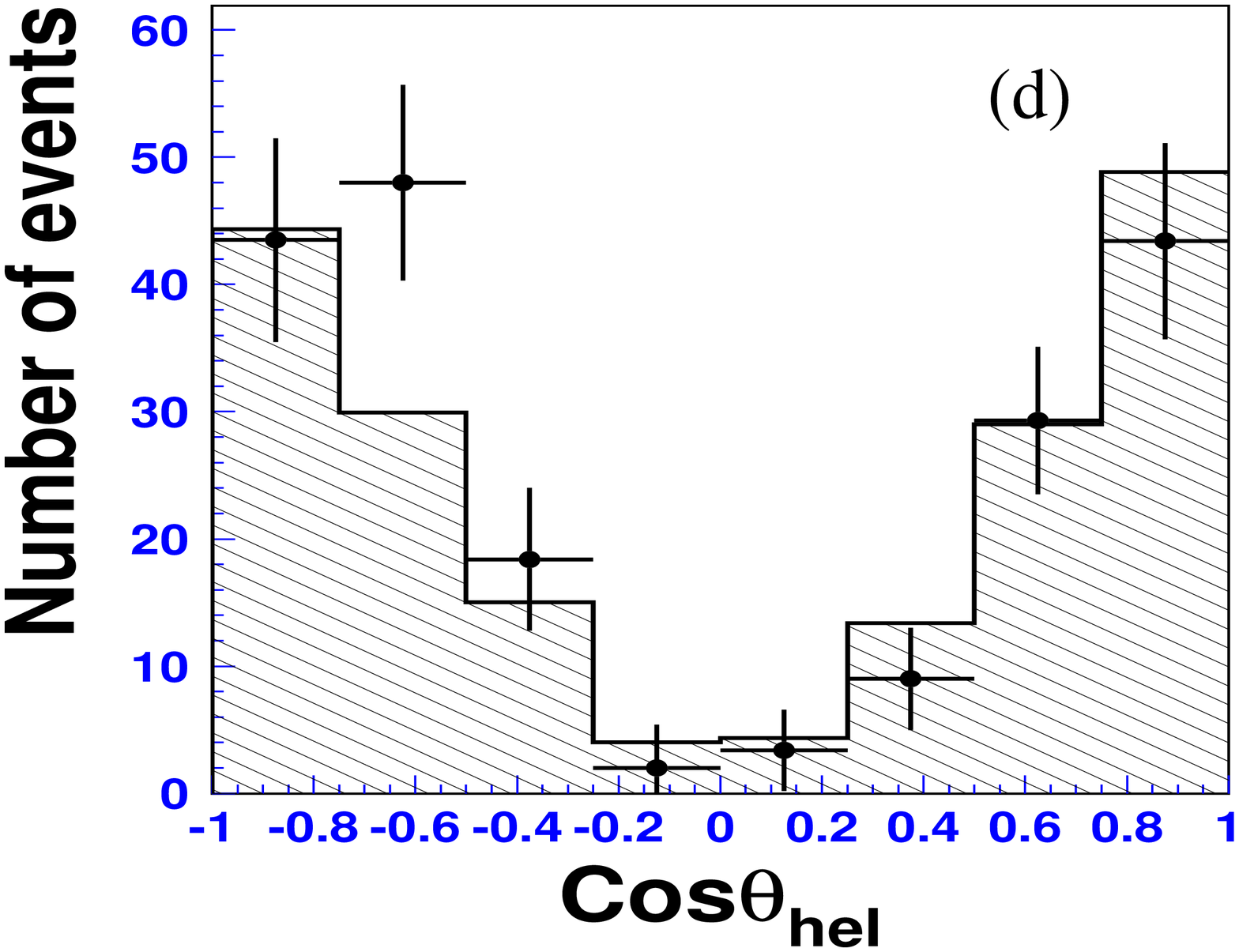,height=3.8cm, width=8.3cm} 
 \end{tabular}
\end{center}
\caption{The distributions in the $B^{-} \rightarrow D^{0}K^{*-}$ signal region for (a) $\Delta E$ with $5.27\,\mathrm{~GeV}/c^2 < M_{\rm{bc}} < 5.29\,\mathrm{~GeV}/c^2$. The solid-line shows the fit, the dashed-line is the signal, the dot-dashed line is coninuum background and the hatched histogram shows the contribution from other $B$ decays. In (b) we show $M_{bc}$ distribution with a $\pm 3\sigma\,\Delta E$ cut where the solid-line shows the fit, the dotted line shows the continuum background and the dashed line is signal. In (c) we show the results of fits to the $\Delta E$ distribution in bins of $K_{S}^{0}\pi^{-}$ invariant mass (points with error bars). In (d) we show the results of fits to the $\Delta E$ distribution in bins of ${\rm{cos}}~\theta_{\rm{hel}}$ (points with error bars). In both (c) and (d) the hatched histogram is a Monte Carlo (MC) simulation of $B^{-} \rightarrow D^{0}K^{*-}$.}
\label{}
\end{figure}

	The signal yields are extracted by a fit to the $\Delta E$ distribution in the region $5.27\,\mathrm{~GeV}/c^2 < M_{\rm{bc}} < 5.29\,\mathrm{~GeV}/c^2$. The signal is parameterized as a Gaussian with parameters determined from MC simulation. The continuum background function is modeled as a first order polynomial function with parameters determined from the $\Delta E$ distribution for the events in the sideband region $5.2\,\mathrm{~GeV}/c^2 < M_{\rm{bc}} < 5.26\,\mathrm{~GeV}/c^2$. Backgrounds from other $B$ decays such as $B^{-} \rightarrow {D^{*}}^{0}K^{*-}$, where $D^{*0} \rightarrow D^{0}\pi^{0}$ and $D^{*0} \rightarrow D^{0}\gamma$, are modeled as a smoothed histogram obtained from the MC simulation. The $\Delta E$ and $M_{\rm{bc}}$ distributions are shown in Figs.1(a) and (b), respectively. We observe a signal of 169.5$\pm$15.4 events with  15.4$\sigma$ statistical significance. The fit results are given in Table I. Statistically significant signals are observed for the $B^- \rightarrow D_{f}K^{*-}$ decay in all three $D^{0}$ decay channels. The final branching fraction is calculated from a weighted average of the results from these three channels. 
	
	We remove the $K_{S}^{0}\pi^{-}$ invariant mass requirement and fit the $\Delta E$ distribution in bins of $K_{S}^{0}\pi^{-}$ invariant mass and plot the signal yield from the fit for each bin. The result is shown in Fig.1(c) and is consistent with a pure $DK^{*-}$ signal MC simulation. We see no indication for non-resonant $D^{0}K_{S}^{0}\pi^{-}$ production. Similarly, we remove the $K^{*-}$ helicity angle requirement and fit the $\Delta E$ distribution in bins of ${\rm{cos}}~\theta_{\rm{hel}}$. We then plot the signal yield from the $\Delta E$ fit for each bin. The result is shown in Fig.1(d) and is consistent with the expectation for a pseudoscalar to pseudoscalar-vector decay. 
	
 	The following sources of systematic error are found to be sizeable: the tracking efficiency~(1\%  per track), $\pi^{0}$ efficiency~(4.8\%), $K_{S}^{0}$ efficiency~(4.5\%), fitting of the $B\bar{B}$ background~(5--7\%) and particle identification~(6--12~\%). Other backgrounds including rare decays that could contribute to the $\Delta E$ signal region, are estimated from the $D^{0}$ sideband data~(1~\%). The uncertainty in the $\Delta E$ signal shape parametrization is determined by varying the mean and width of the signal Gaussian parameters within their errors. The uncertainty from the slope of the background is determined by changing its value by its error. Both of the resulting changes are included in the systematic error from fitting. The combined systematic errors from these sources are $9.5\%$ for $B^{-} \rightarrow D^{0}[K^{-}\pi^{+}]K^{*-}$, $11.8\%$ for $B^{-} \rightarrow D^{0}[K^{-}\pi^{+}\pi^{0}]K^{*-}$ and  $15.5\%$ for $B^{-} \rightarrow D^{0}[K^{-}\pi^{+}\pi^{+}\pi^{-}]K^{*-}$. The systematic error in the final branching fraction is calculated  by weighting according to the efficiency times branching fraction of the three $D^{0}$ decay channels~\cite{PDG}. 

\begin{figure}[ht]
\begin{center}
 \begin{tabular}{ll}
   \epsfig{file=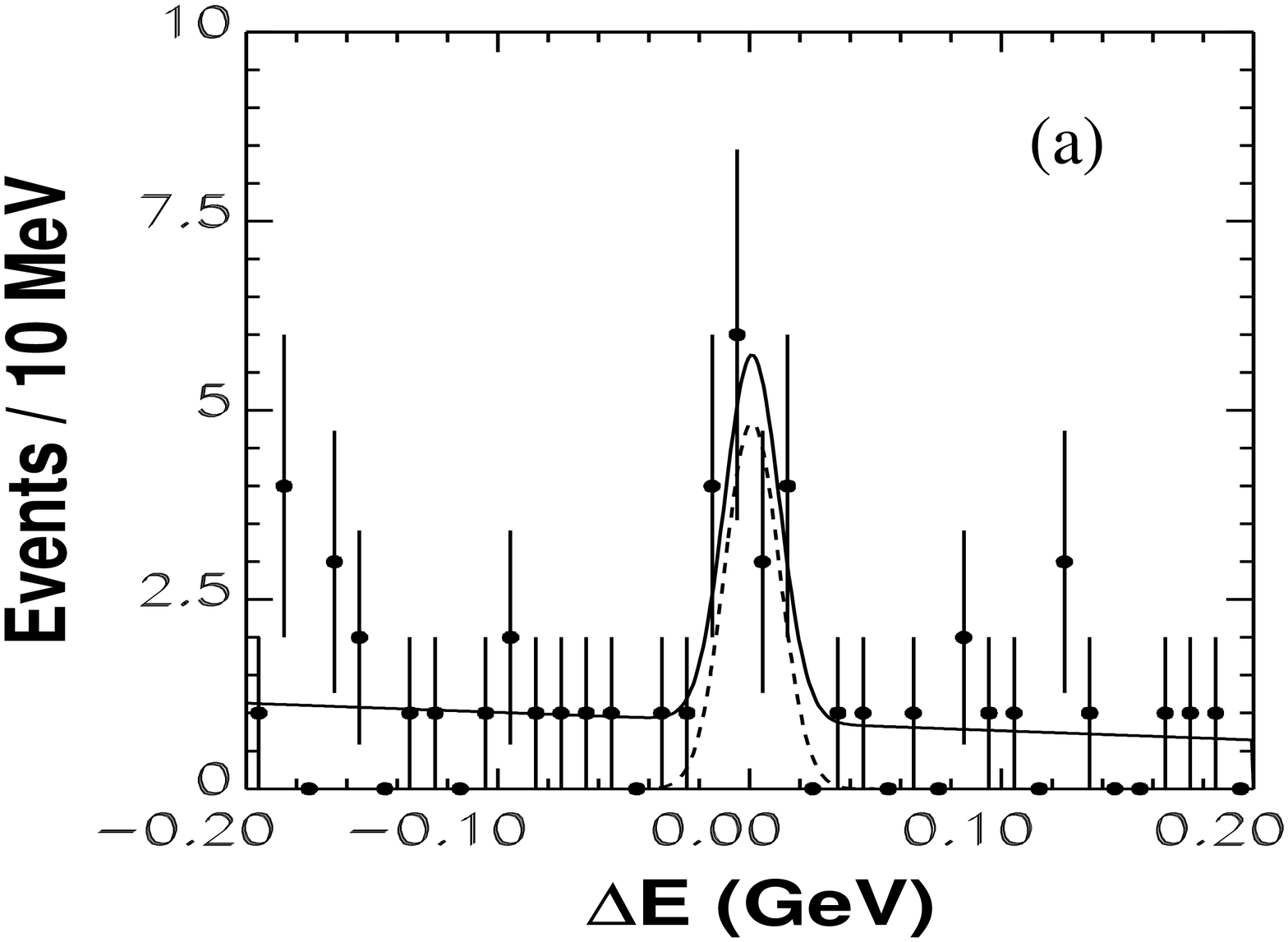 ,height=3.8cm,width=8cm} &
   \epsfig{file=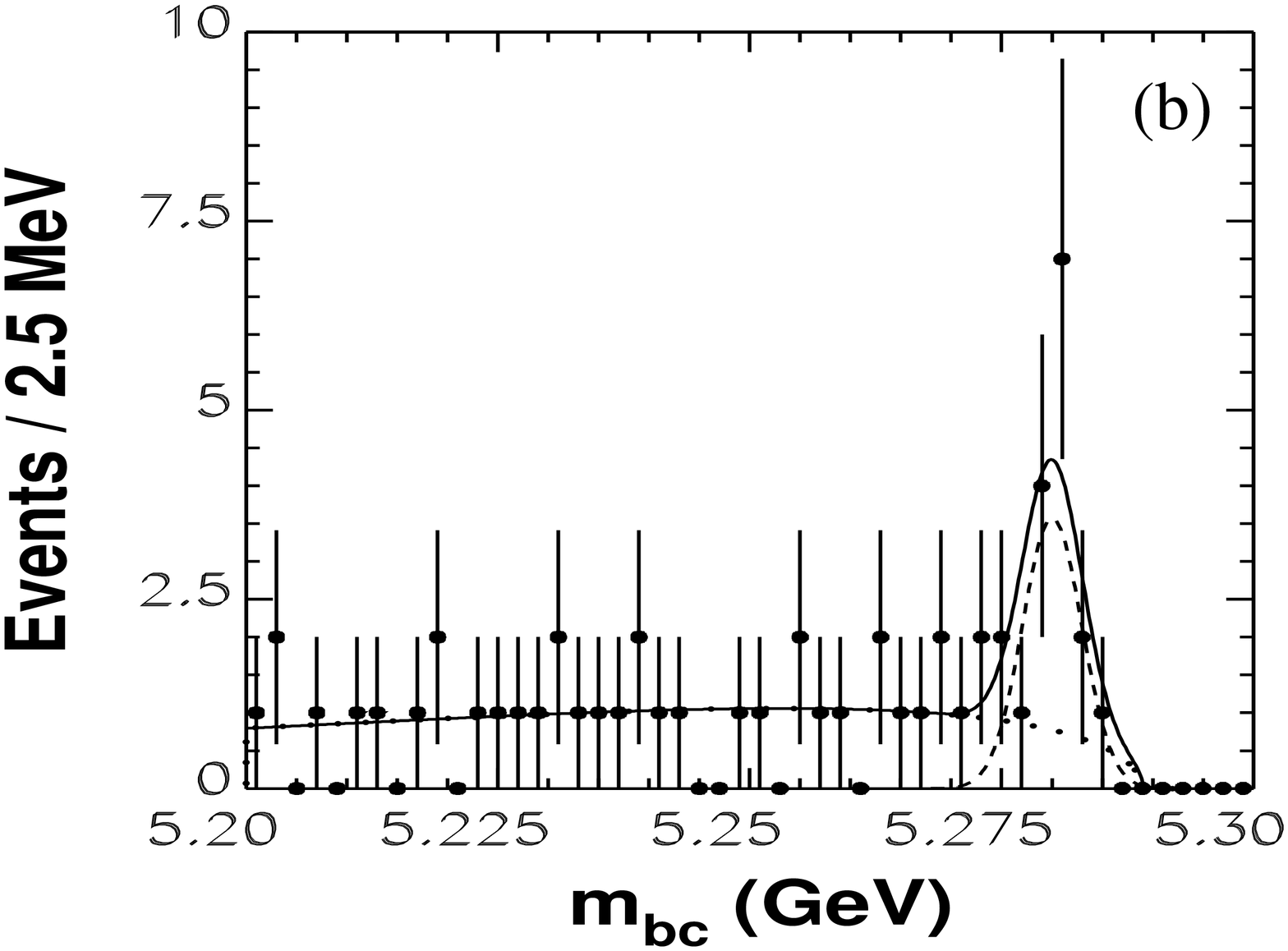 ,height=3.8cm,width=8.0cm}\\ 
   \epsfig{file=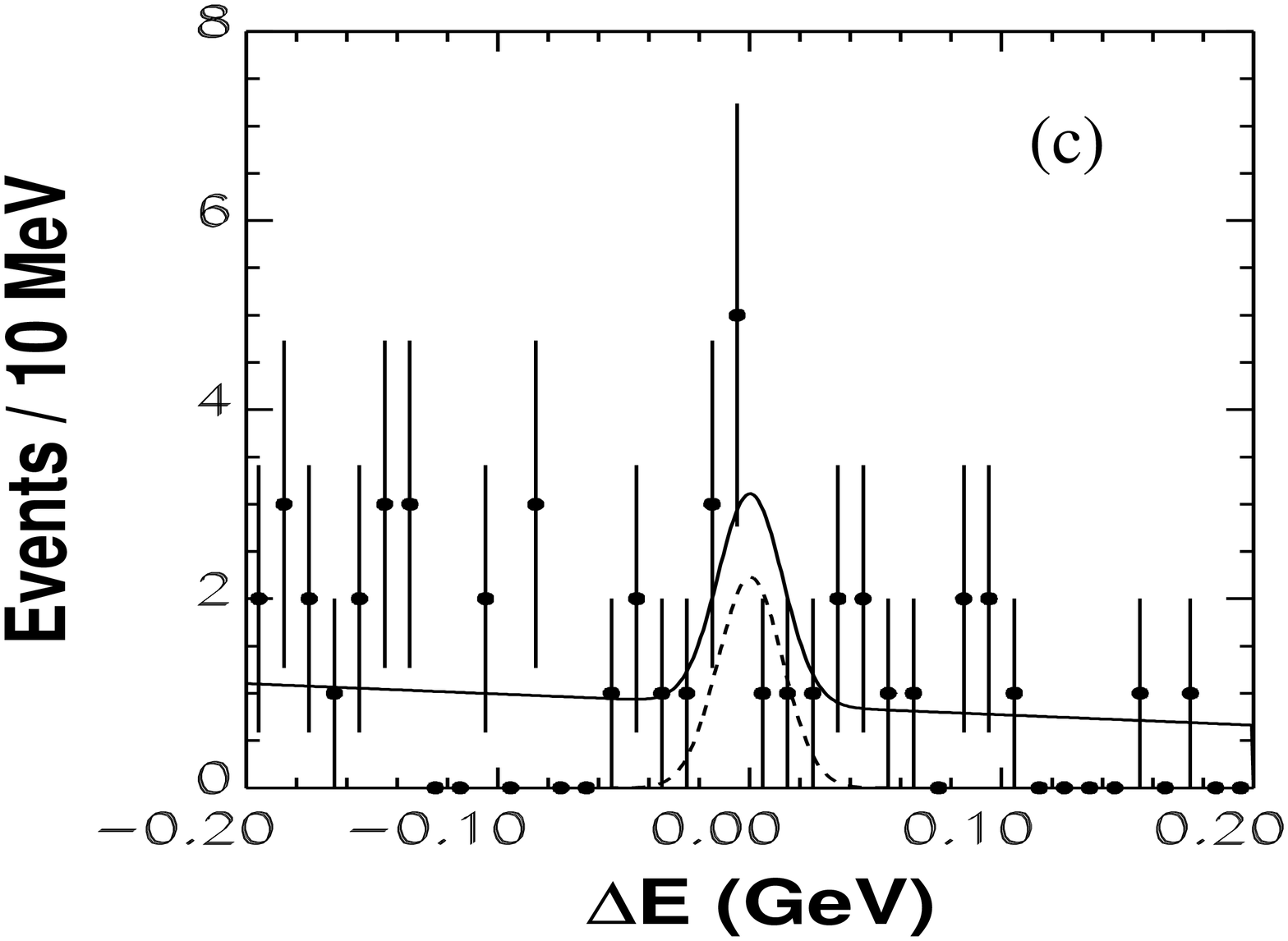 ,height=3.8cm,width=8cm} &
   \epsfig{file=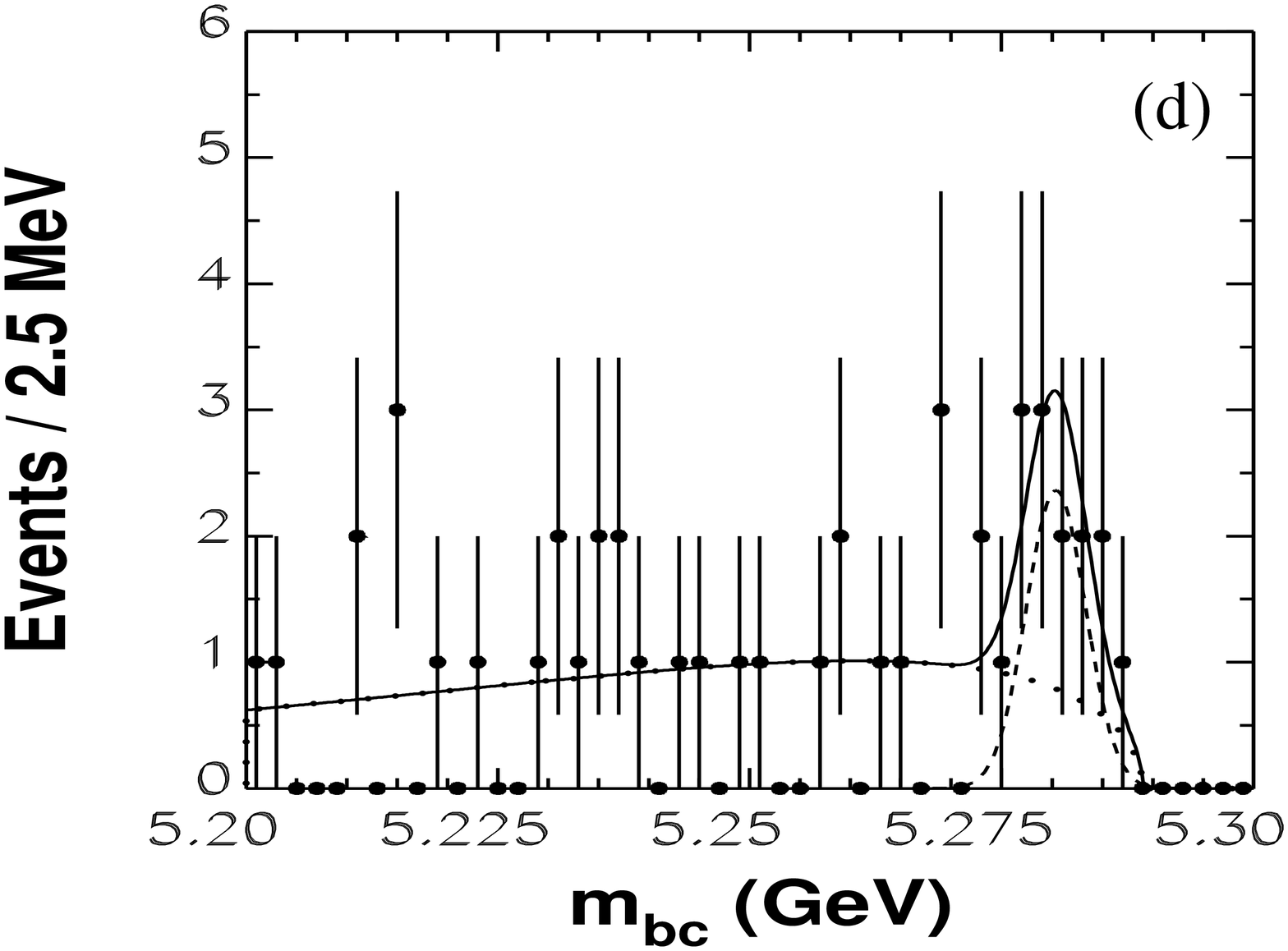 ,height=3.8cm, width=8cm}
 \end{tabular}
\end{center}
\caption{The distributions in the $B^{-} \rightarrow D_{1}K^{*-}$ signal region for (a) $\Delta E$ with $5.27\,\mathrm{~GeV}/c^2 < M_{\rm{bc}} < 5.29\,\mathrm{~GeV}/c^2$ (b)  $M_{bc}$ with a $\pm 3\sigma\,\Delta E$ cut. The distributions in the $B^{-} \rightarrow D_{2}K^{*-}$ signal region for (c) $\Delta E$ with $5.27\,\mathrm{~GeV}/c^2 < M_{\rm{bc}} < 5.29\,\mathrm{~GeV}/c^2$ (d)  $M_{bc}$ with a $\pm 3\sigma\,\Delta E$ cut. The solid-line shows the fit and the dashed-line shows the signal.} 
\label{}
\end{figure}

\begin{table*}[!htb]
\caption{The fit results for $B^- \rightarrow D_{f}K^{*-}$ decay. The signal yields, statistical significance, and branching fraction for each mode are given. The errors shown in the BF are statistical and systematic, respectively.}
\begin{ruledtabular}
\begin{tabular}{lcccc}
decay mode  & Yield($\Delta E$ fit)& Yield($M_{bc}$ fit) & sig & BR($10^{-4}$)  \\ \hline
$D_{f}\rightarrow K^{-}\pi^{+}$ & 67.6$\pm$9.5 & 75.1$\pm$9.5 & 10.8 & 5.0 $\pm$ 0.7 $\pm$ 0.5\\
$D_{f}\rightarrow K^{-}\pi^{+}\pi^{0}$ & 66.3$\pm$9.6  & 65.3$ \pm$9.4 & 9.5 & 6.3 $\pm$ 0.9 $\pm$ 0.7\\
$D_{f}\rightarrow K^{-}\pi^{+}\pi^{+}\pi^{-}$ & 40.3$\pm$7.9 & 43.0$\pm$8.1 & 6.9 & 4.3$\pm$ 0.8 $\pm$ 0.7\\ \hline  
weighted mean & ~~~&~~~~& &5.2 $\pm$0.5(stat) $\pm$ 0.6(sys)\\ 
\end{tabular}
\end{ruledtabular}
\end{table*}

\begin{table*}[!htb]
\caption{Yields, partial-rate charge asymmetries and $90~\%$ C.L intervals for asymmetries.}
\begin{ruledtabular}
\begin{tabular}{lccccc}
Mode & $N(B^{+})$ & $N(B^{-})$ & $\cal{A_{CP}}$ & $90~\%$ C.L   \\ \hline
$ B^\pm \rightarrow D_{f}K^{*\pm}$ & 68.9$\pm$10.1 & 95.3$\pm$11.3 & 0.16 $\pm$ 0.09$\pm$0.08 & $-$0.04$<{\cal A}_f<$0.36\\
$ B^\pm \rightarrow D_{1}K^{*\pm}$ & 6.7$\pm$3.0 & 6.5$\pm$3.1 & $-$0.02 $\pm$ 0.33 $\pm$0.07 & $-$0.57$<{\cal A}_1<$0.53 \\
$ B^\pm \rightarrow D_{2}K^{*\pm}$ & 2.9$\pm$2.4 & 4.3$\pm$2.7 & 0.19 $\pm$ 0.50$\pm$0.04  & $-$0.63$<{\cal A}_2<$1.00 \\
\end{tabular}
\end{ruledtabular}
\end{table*}

	For $B^- \rightarrow D_{CP}K^{*-}$ decay the signal yields are extracted by a fit to the $\Delta E$ distribution in the region $5.27\,\mathrm{~GeV}/c^2 < M_{\rm{bc}} < 5.29\,\mathrm{~GeV}/c^2$ where  $-0.2\,\mathrm{~GeV}<\Delta E<-0.1\,\mathrm{~GeV}$ is excluded. The signal is parameterized as a Gaussian with parameters determined from MC simulation. The continuum background function is modeled as a first order polynomial function with parameters determined from the $\Delta E$ distribution for the events in the sideband region $5.2\,\mathrm{~GeV}/c^2 < M_{\rm{bc}} < 5.26\,\mathrm{~GeV}/c^2$. The $\Delta E$ and $M_{\rm{bc}}$ distributions are shown in Fig.2. We observe a signal of 13.1$\pm$4.3 events for $B^- \rightarrow D_{1}K^{*-}$ and 7.2$\pm$3.6 events for $B^- \rightarrow D_{2}K^{*-}$ with 4.3$\sigma$ and 2.4$\sigma$ statistical significances, respectively. The partial-rate asymmetries ${\cal A}_{1,2}$ are evaluated using signal yields obtained from separate fits to the $B^{+}$ and $B^{-}$ samples. The results are given in Table II.  The systematic uncertainty is from the intrinsic detector charge asymmetry~(1~\%), the $B^-$ and $B^+$ yield extractions~(4--7~\%), and the asymmetry in particle identification efficiency of pions~(1~\%). The systematic error from yield extraction is calculated by changing the fitting parameters by $\pm 1\sigma$. At $90~\%$ C.L intervals, we find
\begin{eqnarray*}
-0.57<{\cal A}_1<0.53, \\
-0.63<{\cal A}_2<1.00~
\end{eqnarray*}
	
In summary, using $88~{\rm fb}^{-1}$ of data collected with the Belle detector, we report a measurement of the exclusive decay mode $B^{-} \rightarrow D_{f}K^{*-}$. This mode has been observed previously~\cite{CLEO}. In this paper, we report a new measurement of the branching fraction ${\cal B}(B^{-} \rightarrow D_{f}K^{*-})$. We also report the partial-rate charge asymmetries for the decay $B^{-} \rightarrow D_{CP}K^{*-}$, where $D_{CP}$ are the neutral $D$ meson CP eigenstates. The measured partial-rate charge asymmetries ${\cal{A}}_{1,2}$ are consistent with zero.

  We wish to thank the KEKB accelerator group for the excellent operation of the KEKB accelerator. We acknowledge support from the Ministry of Education, Culture, Sports, Science, and Technology of Japan and the Japan Society for the Promotion of Science; the Australian Research Council and the Australian Department of Industry, Science and Resources; the National Science Foundation of China under contract No.~10175071; the Department of Science and Technology of India; the BK21 program of the Ministry of Education of Korea and the CHEP SRC program of the Korea Science and Engineering Foundation; the Polish State Committee for Scientific Research under contract No.~2P03B 17017; the Ministry of Science and Technology of the Russian Federation; the Ministry of Education, Science and Sport of the Republic of Slovenia; the National Science Council and the Ministry of Education of Taiwan; and the U.S.\ Department of Energy.


\begin{thebibliography}{99}
\bibitem{gamma} Another naming convention, 
$\gamma$ ($= \phi_3$), is also used in the literature.
\bibitem{CC}
Throughout this paper, the inclusion of the charge conjugate mode decay is implied
unless otherwise stated.
\end{thebibliography}
\end{document}